\documentclass[amssymb,aps,floatfix]{revtex4}
\emergencystretch=\maxdimen
\usepackage{graphicx}
\usepackage{dcolumn}
\usepackage{amsmath}
\usepackage[latin1]{inputenc}
\usepackage{amssymb}
\usepackage[colorlinks=true, citecolor=blue, urlcolor = blue, linkcolor= red, bookmarks=true]{hyperref}
\usepackage{float}
\usepackage{amsfonts}
\usepackage{dcolumn}
\usepackage{hyperref}
\usepackage{subfigure}
\usepackage{pgfplots}
\usepackage{booktabs}
\usepackage{mathrsfs}
\usepackage{multirow}
\usepackage{afterpage}
\pgfplotsset{compat=1.16}

\usepackage{wrapfig}
\usepackage{adjustbox}
\makeatletter
\def\btt#1{\texttt{\@backslashchar#1}}
\DeclareRobustCommand\bblash{\btt{\@backslashchar}} \makeatother

\begin{document}

\title[]{Strong gravitational lensing by rotating Simpson--Visser black holes}
\author{Shafqat Ul Islam$^{a}$ } \email{shafphy@gmail.com}
\author{Jitendra Kumar$^{a}$ } \email{jitendra0158@gmail.com}
\author{Sushant~G.~Ghosh$^{a}$} \email{sghosh2@jmi.ac.in, sgghosh@gmail.com}

\affiliation{$^{a}$ Centre for Theoretical Physics, 
	Jamia Millia Islamia, New Delhi 110025, India}

\begin{abstract}
We investigate strong field gravitational lensing by rotating Simpson-Visser black hole, which has an additional parameter ($0\leq l/2M \leq1$), apart from mass ($M$) and rotation parameter ($a$). A rotating Simpson-Visser metric correspond to (i) a Schwarzschild metric for $l/2M=a/2M=0$ and $ M \neq 0 $, (ii) a Kerr metric for $l/2M=0$,   $|a/2M|< 0.5$ and $ M \neq 0 $  (iii) a rotating regular black hole metric for  $|a/2M|< 0.5$, $ M \neq 0 $ and $l/2M$ in the range  $0<l/2M<0.5 + \sqrt{\left(0.5\right)^2-(a/2M)^2}$, and (iv) a  traversable wormhole for a  $|a/2M|  >0.5$ and $l/2M\neq 0$. We find a decrease in the deflection angle $\alpha_D$ and also in the ratio of the flux of the first image and all other images  $ r_{mag}$.   On the other hand, angular position $\theta_{1}$ increases more slowly and  photon sphere radius $x_{m}$ decreases more quickly, but angular separation $s$ increases more rapidly, and their behaviour is similar to that of the Kerr black hole. The formalism is applied to discuss the astrophysical consequences in the supermassive black holes NGC 4649,  NGC 1332, Sgr A* and M87* and find that the rotating Simpson-Visser black holes can be quantitatively distinguished from the Kerr black hole via gravitational lensing effects. We find that the deviation of the lensing observables $\Delta\theta_1$ and $\Delta s$ of rotating Simpson Visser black holes from Kerr black hole for $0<l/2M <0.6$ ($a/2M=0.45$),  for supermassive black holes Sgr A* and M87, respectively, are in the range  $0.0422-0.11658~\mu$as and $0.031709-0.08758~\mu$as  while $|\Delta r_{mag}|$ is in the range $0.2037 - 0.95668$.   It is difficult to distinguish the two black holes because the departure are in $\mathcal{O}(\mu$as), which are unlikely to get resolved by the current EHT observations, and one has to wait for future observations by ngEHT can pin down the exact constraint.
We also derive a two-dimensional lens equation and formula for deflection angle in the strong field limit by focusing on trajectories close to the equatorial plane.
\end{abstract}

\maketitle
\thispagestyle{empty}
\setcounter{page}{1}

\section{Introduction}
That the end state of gravitational collapse of a sufficiently massive star ($\sim3.5M_{\odot}$) is a gravitational singularity is a fact established by the singularity theorems of Hawking and Penrose \cite{Hawking:1970}  (see also Ref. \cite{Hawking:1973}). The existence of a singularity, by its very definition, means spacetime ceases to exist, marking a breakdown of the laws of physics.  The extreme condition, in any form, that may exist at the singularity implies that one should rely on quantum gravity which is expected to resolve this singularity \cite{Wheeler:1964}. In the absence of any definite quantum gravity, which can allow one to understand the black hole interior and resolve the singularity issue, significant attention shifted to regular models. The idea of regular models was pioneered by Sakharov \cite{Sakharov:1966} and Gliner \cite{Gliner:1966}, that suggests one can get rid of the singularities by considering a matter, i.e., with a de Sitter core, with the equation of state $p=-\rho$. The cosmological vacuum obeys this equation of state, and hence, $T_{\mu \nu}=\Lambda g_{\mu \nu}$; $\Lambda$ is the cosmological constant.

Bardeen \cite{Bardeen:1968}, motivated by  Sakharov \cite{Sakharov:1966} and Gliner \cite{Gliner:1966}, proposed the first regular black hole  with horizons, but there is no central singularity \cite{Ansoldi:2008jw,Ghosh:2014pba}. The Bardeen black hole near the origin acts like the de Sitter spacetime, whereas for large  $r$, it is according to Schwarzschild black hole \cite{Ansoldi:2008jw}. Thus the black hole interior does not result in a singularity but develops a de Sitter like region, eventually settling with a regular center. Thus its maximal extension is the one of the Reissner\(-\)Nordstr$\mathrm{\ddot{o}}$m spacetime, but with a regular center \cite{Borde:1994ai,Borde:1996df}.  Later, Ayon-Beato and Garcia \cite{Ayon:1999}  demonstrate that the Bardeen's model is an exact solution of general relativity coupled to nonlinear electrodynamics thereby an alteration of the Reissner\(-\)Nordstr$\mathrm{\ddot{o}}$m black hole solution \cite{Reissner:1916}.  There has been an enormous advance in the analysis and application of regular black holes,  \cite{AyonBeato:1998ub,Lemos:2011dq,Cisterna:2020rkc}  and also on regular rotating black \cite{Ghosh:2014pba,Kumar:2020cve,Eichhorn:2021etc} holes. Most of these solutions are, more or less, based on Bardeen's proposal that has non-linear electrodynamics as the source.    
It turns out that most regular  black holes have a core that is asymptotical de Sitter (with constant positive curvature) \cite{Bardeen:1968,AyonBeato:1998ub}. One exception is the regular black hole in Refs \cite{hc,Ghosh:2014pba,Simpson:2019mud} has an asymptotically Minkowski core \cite{Simpson:2019mud} thereby greatly simplifying the physics in the deep core. 

Lately, another interesting spherically symmetric regular black hole spacetime is proposed by Simpson-Visser \cite{Simpson:2018tsi,Simpson:2019cer,Lobo:2020ffi} given by
\begin{equation} \label{sv}
	ds^{2}=-\left(1-\frac{2m}{\sqrt{r^{2}+l^{2}}}\right)dt^{2}+\frac{dr^{2}}{1-\frac{2m}{\sqrt{r^{2}+l^{2}}}}
	+\left(r^{2}+l^{2}\right)\left(d\theta^{2}+\sin^{2}\theta \;d\phi^{2}\right).
\end{equation}
where $l$ is a constant with $l \neq 0$ yields the least modification of the Schwarzschild black hole spacetime and encompasses it when $l=0$.   The parameter $l$ has dimensions of length introduced to cause a repulsive force to avoid singularity and is responsible for the regularisation of the metric at $r=0$. This spacetime (\ref{sv}) interpolates between the  Schwarzschild black hole and the Morris--Thorne traversable wormhole. The Simpson-Visser metric  is regular everywhere;  it is evident  by analysing the scalar invariant $R_{ab}R^{ab}$ and $R_{abcd}R^{abcd} $ which are well behaved everywhere including at $r=0$ \cite{Simpson:2018tsi}.  The Simpson-Visser spacetime received significant attention starting with a discussion on the energy conditions, causal structure, and innermost stable circular orbit (ISCO)  \cite{Simpson:2018tsi}. Later, a cascade of work followed, viz. a Vaidya radiating spacetime and traversable wormhole \cite{Simpson:2019cer}, regularity, quasi-local mass, energy conditions and the causal structure of black bounce models \cite{Lobo:2020ffi}, gravitational lensing in a strong deflection limit \cite{Tsukamoto:2020bjm}, construction of rotating counterparts  \cite{Mazza:2021rgq}, the slow rotation generalization and properties of the wormhole branch of this solution  \cite{Bronnikov:2021liv}, and discussion on black holes shadows \cite{Junior:2021atr, Shaikh:2021yux}. 

The gravitational lensing, one of the earlier predictions of general relativity, is also considered a test of general relativity, primarily being in the strong field regime\cite{Crispino:2019yew}. Since then, gravitational lensing proved to be an essential tool for providing insights on the spacetime \cite{Einstein:1936yew}. In the strong field scenario, the light passes very close to the black hole such that the bending angle is much greater than 1 radian and photon can encircle the black hole more than once which results in shadow~\cite{Synge:1966}, photon rings \cite{Gralla:2019xty}, relativistic images \cite{Darwin:1959,Bozza:2009yw,Bozza:2001xd,Bozza:2002zj,Bozza:2002af}. Although relativistic images can not be resolved due to their small separation \cite{Virbhadra:1998dy,Petters:2002fa} and thin width \cite{Gralla:2019xty}, but with the help of the new generation Event Horizon Telescope (ngEHT), they would be accessible and help in distinguishing different kinds of black holes, thus providing a precise test in the strong field limit.
  
 Lately, there has been an enormous discussion of gravitational lensing from the strong field by black holes \cite{Virbhadra:1999nm,Frittelli:1999yf,Bozza:2001xd,Bozza:2002zj,Bozza:2002af,Bozza:2009yw,Eiroa:2002mk,Iyer:2006cn,Tsukamoto:2016jzh,Virbhadra:2007kw,Shaikh:2019jfr}, since it  could provide another avenue to test general relativity.  Virbhadra and Ellis \cite{Virbhadra:1999nm} have given a numerical recipe that allows studying the large deflection of light rays, resulting in strong gravitational lensing. Later, the strong field gravitational lensing for general spherically symmetric and static spacetime analysed analytically by Bozza in \cite{Bozza:2001xd,Bozza:2002zj,Bozza:2002af,Bozza:2009yw} and also by Tsukamoto in \cite{Tsukamoto:2016jzh}. The observationally gravitational lensing by black holes began to be crucial in the 1990s, which triggered quantitative studies of the gravitational lensing by Kerr black holes \cite{Bozza:2009yw,Rauch:1994qd,Bozza:2008mi}. Later, due to the tremendous advancement of current observational facilities,  gravitational deflection of light by rotating black holes has received significant attention \cite{Ghosh:2020spb,Wei:2011nj,Beckwith:2004ae,Hsiao:2019ohy,Kapec:2019hro,Gralla:2019drh,James:2015yla}. In addition, gravitational lensing  by hairy Kerr black holes, compared to Kerr black hole,  exhibit several interesting feature \cite{Cunha:2019ikd}.  The recent time witnessed significant attention in investing strong gravitational lensing by black holes is due to the EHT  results, which released the image of the supermassive  M87*. Most galaxies harbour a supermassive black hole in their central region, like M87*, in the elliptical galaxy and the Milky Way has Sagittarius A* \cite{Ghez:1998ph}. 
 
The goal is to compare the observational predictions of rotating Simpson-Visser black holes with those of the Kerr black hole in an identical observational situation, i.e.,  same source and observer position and same astrophysical black holes.  We only consider an isolated black hole with no surrounding accretion disk or plasma. It is possible that the corrections in the deflection angle,  by rotating Simpson-Visser black holes, in the Kerr black holes can be explained by other environmental effects also, such as disk, plasma, dark matter \cite{Aldi:2016ntn,Bozza:2010xqn}.   A study of deflection of light by Schwarzschild and Kerr black hole in the presence of medium has been presented both in weak and strong field regimes \cite{Perlick:2000,Tsupko:2013cqa,Bisnovatyi}.  Interestingly, the plasma environment around the black hole considerably changes the position and magnification of the relativistic images  \cite{ZamanBabar:2021aqv}. The presence of the uniform plasma around the black hole increases the angular position of the images, and their magnification \cite{Tsupko:2013cqa,Liu:2016eju}.

 This paper investigates the strong gravitational lensing by rotating Simpson-Visser black holes  \cite{Mazza:2021rgq} and assess the phenomenological differences with the Kerr black holes; in particular, we discuss the effect of deformation parameter $l$  on gravitational lensing observables and time delay between the relativistic images.  Further, considering the supermassive black holes, NGC 4649,  NGC 1332, Sgr A* and M87* as the lens, we obtain the positions, separation, magnification, and time delay in the formation of relativistic images.  We have also discuss quasi-equatorial gravitational lensing by rotating Simpson-Visser black holes  in the strong field limit. Besides the spin, we also expect to study the influence
of the deviation parameter $l$ on black hole lensing.
 
The paper is organized as follows: In the Sec. \ref{Sec2}, we briefly review the recently obtained rotating Simpson-Visser black holes  \cite{Mazza:2021rgq} and also discuss its horizon structure to put restrictions on parameters and in turn discuss the gravitational deflection of light in the strong field limit for rotating Simpson-Visser black holes  \cite{Mazza:2021rgq} which is the subject in Sect. \ref{Sec3}.  In Sec. \ref{Sec3},  we also discuss the strong lensing observables by rotating Simpson-Visser black holes, including the image positions $\theta_{\infty}$, separation $s$, magnifications  and time delays  of the images. By taking the supermassive black holes NGC 4649,  NGC 1332, Sgr A* and M87*, as the lens, we  numerically estimate the observables in Sec. \ref{Sec5}. In Sec. \ref{Sec7} we discuss the non-equatorial
gravitational lensing and compute the deflection angle in the strong field limit and finally, we summarize our results and evoke some perspectives to end the paper in Sect. \ref{Sec6}. 

 Throughout this paper, unless otherwise stated, we adopt natural units ($8 \pi G\; =\; c\;=\; 1$)

\section{Rotating Simpson-Visser black holes}\label{Sec2}
The metric of the rotating Simpson-Visser black holes is constructed as a modification from the metric of the Kerr black hole with additional parameter $l$ apart from mass $M$ and angular momentum $a$, which  in Boyer-Lindquist coordinates, reads as \cite{Mazza:2021rgq} 
\begin{eqnarray}\label{metric1}
ds^2 &=& -\left(1-\frac{2M \sqrt{r^2+l^2}}{\Sigma}\right) dt^2 + \frac{\Sigma}{\Delta} dr^2+ \Sigma d\theta^2 -\frac{4Ma \sin^2\theta\sqrt{r^2+l^2}}{\Sigma} dtd\phi \nonumber \\ && + \frac{\mathbb{A}\sin^2\theta~}{\Sigma} d\phi^2,
\end{eqnarray}
where $\Sigma = r^2+l^2 +a^2\cos^2\theta$,  $\Delta = r^2+l^2+a^2-2M\sqrt{r^2+l^2}$, and $\mathbb{A} =(r^2+l^2 +a^2)^2-\Delta a^2\sin^2\theta$. The parameter $l$ describes deviation from Kerr black hole and one finds that Kerr black hole is encompassed as a special case ($l=0$). When $a =0$, the above metric reduces to the spherical Simpson--Visser metric~(\ref{sv})~\cite{Simpson:2018tsi,Simpson:2019cer,Lobo:2020ffi}.  When both the parameters $a$ and $l$ are zero  it gives the familiar Schwarzschild solution. The non-negative parameter $l$ with dimensions of length is introduced to cause a repulsive force which avoids singularity at $r=0$. The metric~(\ref{metric1}), is symmetric under the transformation $r\to-r$, and is thus composed of identical portions around $r=0$. The surface $r=0$, is a regular surface of finite size for $l\ne 0$, which the observer can easily cross~\cite{Mazza:2021rgq}. Unlike Bardeen \cite{Bardeen:1968}  or Hayward spacetimes, metric~(\ref{metric1}) is either a regular black hole or a traversable worm-hole, depending on the value of the parameter $l$~\cite{Mazza:2021rgq}.  

\begin{figure*}
	\begin{centering}
		\begin{tabular}{c c}
		    \includegraphics[scale=0.75]{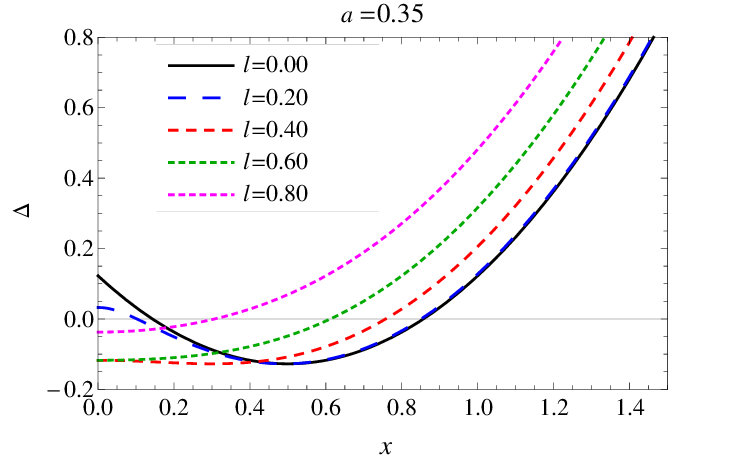}&
		    \includegraphics[scale=0.75]{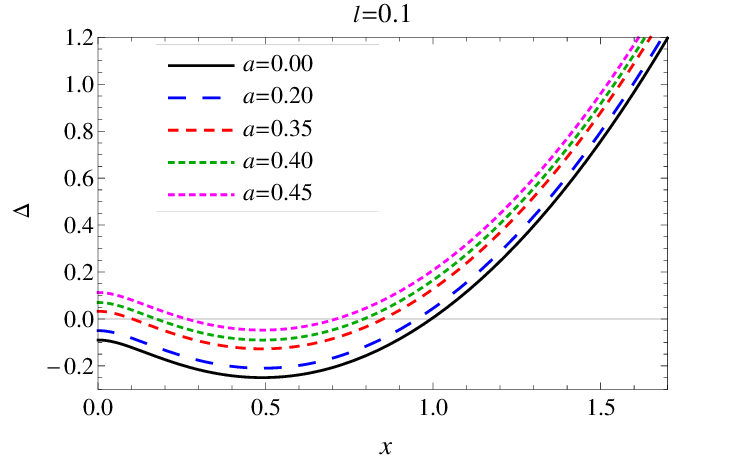}\\
		    \includegraphics[scale=0.75]{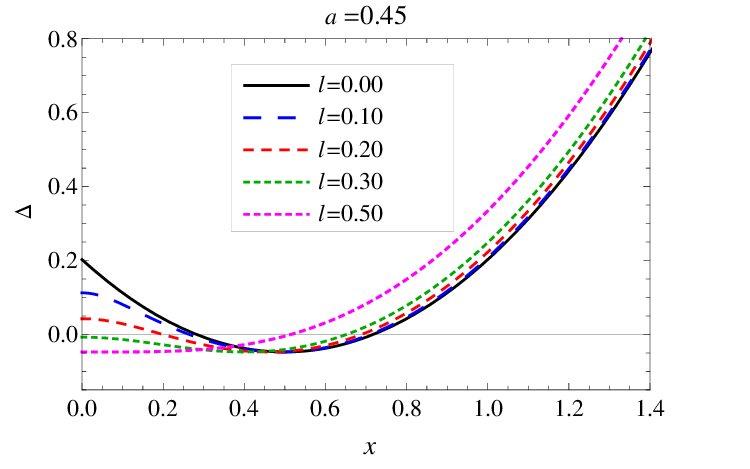}&
		    \includegraphics[scale=0.75]{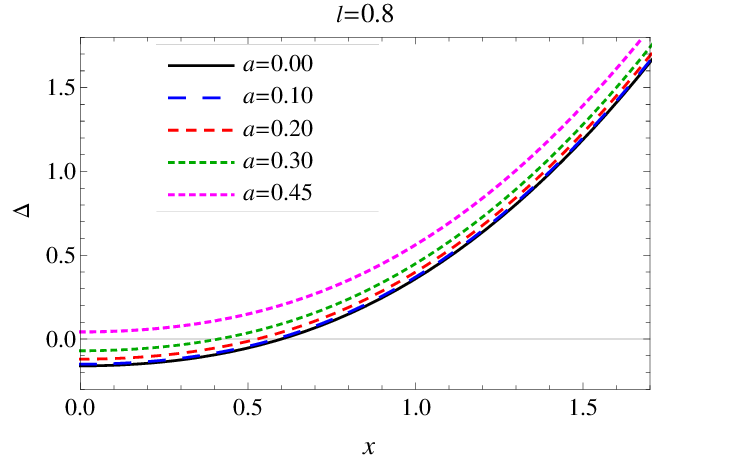}
			\end{tabular}
	\end{centering}
	\caption{The horizons (zeroes of $\Delta$ ) of rotating Simpson-Visser black holes. We restrict the parameter $l$ in the range $0 < l < 0.5+ \sqrt{\left(0.5\right)^2-a^2}$ as we consider only black holes. The case $l=0$ (Kerr black hole) is shown for comparison. For $l>l_{c}$ (e.g., for $a=0.35$, $l_{c} \approx 0.1429$ and for $a=0.45$, $l_{c} \approx 0.282$), the black holes admit only event horizon. We  measure all the quantities in units of Schwarzschild radius $2M$. }\label{plot1}	
\end{figure*}

We  measure  the quantities $r, a, t$,  and $l$ in units of Schwarzschild radius $2M$ \cite{Bozza:2002zj} and use $x$ instead of radius $r$  to rewrite the metric~(\ref{metric1}) as

\begin{eqnarray}\label{metric2}
ds^2 &=& -\left(1-\frac{ \sqrt{x^2+l^2}}{\Sigma}\right) dt^2 + \frac{\Sigma}{\Delta} dx^2+ \Sigma d\theta^2 -\frac{2a \sin^2\theta\sqrt{x^2+l^2}}{\Sigma} dtd\phi  + \frac{\mathbb{A}\sin^2\theta~}{\Sigma} d\phi^2,
\end{eqnarray}
where $\Sigma = x^2+l^2 +a^2\cos^2\theta$, $\Delta = x^2+l^2+a^2-2\sqrt{x^2+l^2}$ and $\mathbb{A} =(x^2+l^2 +a^2)^2-\Delta a^2\sin^2\theta$.
\begin{figure}[t]
	\begin{centering}
		\begin{tabular}{c c}
		    \includegraphics[scale=0.85]{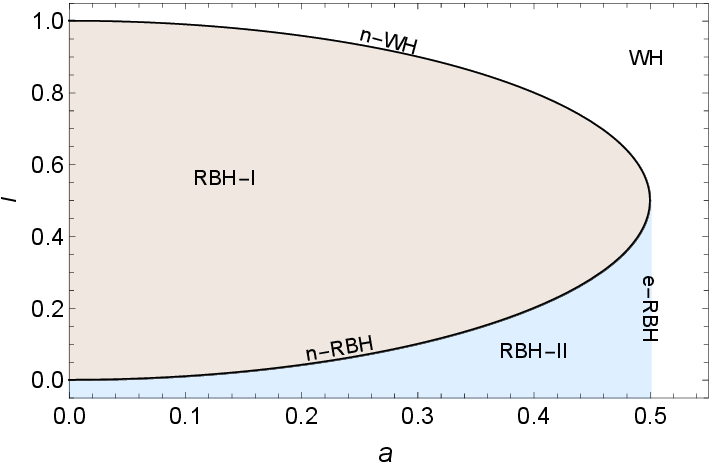}&
		    \end{tabular}
	\end{centering}
	\caption{Parameter space for the rotating Simpson-Visser black hole and the corresponding spacetime structure. For $a=0.3$  the rotating Simpson-Visser black hole admits  both Cauchy and  event horizon for $0.0<l_{cb}<0.1$ and only event horizon for $0.1<l_{c}<0.9$.  For $a=0.45$,  both Cauchy and  event horizon exist for $0.0<l_{cb}<0.282055$  and only event horizon is present for $0.282055<l_{c}<0.717945$.  We  measure  the quantities $a$ and $l$ in units of Schwarzschild radius $2M$.}\label{plota}		
\end{figure}  
Interestingly, there  exist values of $a$ and $l$ for which
$\Delta=0$, which correspond to the coordinate singularities. The points of coordinate singularities, if they exist, are also the horizons of the black hole  metric (\ref{metric2}). Thus, the horizons are located at
\begin{equation}\label{ksh}
x_{\pm} = \left[\left(0.5 \pm \sqrt{\left(0.5\right)^2-a^2}\right)^2-l^2\right]^{1/2} ,
\end{equation}
 The horizons of the metric~(\ref{metric2}) exist when $|a|<0.5$ and $l<0.5+\sqrt{\left(0.5\right)^2-a^2}$. For $l < 0.5 - \sqrt{\left(0.5\right)^2-a^2}$, we have two horizons, $x_{\pm}$, corresponding to the radii of outer (event) ($x_{+}$) and inner (Cauchy) ($x_{-}$) horizons  (This corresponds to RBH-II in Fig~\ref{plota}). Whereas, when $0.5-\sqrt{\left(0.5\right)^2-a^2} < l < 0.5 + \sqrt{\left(0.5\right)^2-a^2} $, only event horizon exists  (This corresponds to RBH-I in Fig~\ref{plota}, also cf. Fig.~\ref{plot1}).  For $|a|=0.5$ and $l < 0.5$, Cauchy and event horizon correspond to extremal black hole (e-RBH) with degenerate horizons, while for $|a|=0.5$ and $l > 0.5$, there are no horizons and so no black hole. The case $|a| > 0.5$~ with no horizons, the metric~(\ref{metric2}) describes a traversable wormhole with a time-like throat, which becomes null for $|a| = 0.5$ \cite{Mazza:2021rgq}.

We shall be considering only black holes and restrict the values of $a$ and $l$ in the range $|a|<0.5$ and $0<l<0.5 + \sqrt{\left(0.5\right)^2-a^2}$ for at least one horizon to exist to qualify the spacetime as a black hole. As the solution is symmetric around $x=0$, the light deflection will be the same on either side. Without loss of generality, we shall consider $x>0$. Images are formed on both sides of the lens.  We have to carefully choose the correct sign for the angular momentum as all strong field limit coefficients depend on $a$. Conventionally the direction of the black hole spin is in the north. The photons winding counterclockwise are described by a positive $a$ and form images on the eastern side of the black hole. Images formed by retrograde rays are described taking a negative $a$ and appear on the western side.
 
The timelike killing vector $\xi^a =(\frac{\partial}{\partial t})^a$
of the rotating Simpson-Visser metric has norm
\begin{equation}\label{tls}
\xi^a \xi_a = g_{tt} = 1-\frac{1}{\sqrt{\Sigma}},
\end{equation}
becomes positive in the region where $ x^2+l^2+a^2\cos^2\theta- \sqrt{x^2+l^2} < 0.$
For the rotating Simpson-Visser metric, there also exists the static limit surface (SLS) that can be determined by  the zeros $ g_{tt}(x)=0 $, i.e., of  $ x^2+l^2+a^2\cos^2\theta- \sqrt{x^2+l^2} = 0. $
The region between event horizon and SLS is called ergosphere, where the asymptotic time translation Killing field $\xi^{t}$ a becomes spacelike and an observer follow an orbit of $\xi^{t}$. 

\section{Gravitational lensing in strong deflection limit }\label{Sec3}
The rotating Simpson-Visser metric like the Kerr metric has reflection symmetry $\theta \to \pi - \theta$, due to which the light rays which are initially in the equatorial plane will always remain in the equatorial plane, making it easier to consider the motion of light rays in this plane. Further, we consider the source to be almost aligned along the optical axis, and the source is located behind the lens, which allows the images to be very prominent and highly magnified. 

Next, we consider  the lens equation given in \cite{Bozza:2002zj,Bozza:2008ev},   
\begin{eqnarray}\label{lenseq}
\beta &=& \theta -\frac{D_{LS}}{D_{OL}+D_{LS}} \Delta\alpha _n,
\end{eqnarray}  
where $\Delta\alpha _n = \alpha_{D}(\theta)$ - $2n\pi $, is the extra deflection angle  with  $n \in N $ and $ 0 < \Delta\alpha _n \ll 1 $. Here, $\alpha_{D}(\theta)$, $\beta$ and $\theta$ are, respectively, the deflection angle, angular separation of source and image from optical axis.  The distance of source, lens from the observer are given by $D_{OS}$ and $D_{OL}$. The  lens equation have been used in more general ways \cite{Perlick:2003vg,Eiroa:2003jf,Frittelli:1998hr,Frittelli:1999yf} but the lens Equation~(\ref{lenseq})  precisely helps to analyse the observational effects. The rotating Simpson-Visser black hole in the equatorial plane ($\theta= \pi/2$) takes the form
\begin{eqnarray}\label{NSR}
\mathrm{ds^2}=-A(x)dt^2+B(x) dx^2 +C(x)d\phi^2-D(x)dt\,d\phi,
\end{eqnarray}
where
\begin{equation}\label{compo}
A(x)= 1-\frac{1}{\sqrt{\Sigma}},~~~~
B(x)=\frac{\Sigma}{\Delta}\nonumber,~~~~
C(x)= \frac{\mathbb{A}}{\Sigma},~~~~
D(x)=\frac{2 a }{\sqrt{\Sigma}},    
\end{equation}
and $\Sigma = x^2+l^2 $ and $\mathbb{A} =(x^2+l^2 +a^2)^2-\Delta a^2$. Using the null geodesic equations which are  first order ordinary differential equations \cite{Chander:1992pc}, the deflection angle of a photon moving on the equatorial plane for rotating Simpson-Visser black hole reads \cite{Bozza:2002af}
\begin{eqnarray}\label{bending2}
\alpha_{D}(x_0)=2\int_{x_0}^{\infty}\frac{d\phi}{dx}dx = 2\int_{x_0}^{\infty}\frac{\sqrt{A_0 B }\left(2AL+ D\right)}{\sqrt{4AC+D^2}\sqrt{A_0 C-A C_0+L\left(AD_0-A_0D\right)}} dx-\pi,
\end{eqnarray}
where $A_0$ is the value of $A(x)$ at $x=x_0$, which is the light's closest approach. The quantities $C_0$ and $D_0$ are defined in the similar way as $A_0$ and $L$  is the projection of  angular momentum.  Unlike in weak deflection limit, where the deflection angle is minimal and an approximate form can be obtained, in the case of strong deflection limit (SDL), the argument of the integral~(\ref{bending2}) is expanded near the photon sphere. The photon sphere radius $x_m$ corresponds to the critical impact parameter and  is the solution  of the  following equation~\cite{Bozza:2002af}  
\begin{eqnarray}\label{ps}
A(x)C'(x)-A'(x)C(x) + L(A'(x)D(x) - A(x)D'(x)) &=& 0,
\end{eqnarray}
which implies 
\begin{eqnarray}\label{ps1}
x \sqrt{l^2+x^2} \left(2 a \sqrt{a^2-\sqrt{l^2+x^2}+l^2+x^2}-2 a^2-5 l^2-5 x^2\right)+x \left(4 l^2 x^2+2 l^4+3 l^2+2 x^4+3 x^2\right)=0
\end{eqnarray}

\begin{figure}[t]
	\begin{centering}
		\begin{tabular}{c c}
		    \includegraphics[scale=0.77]{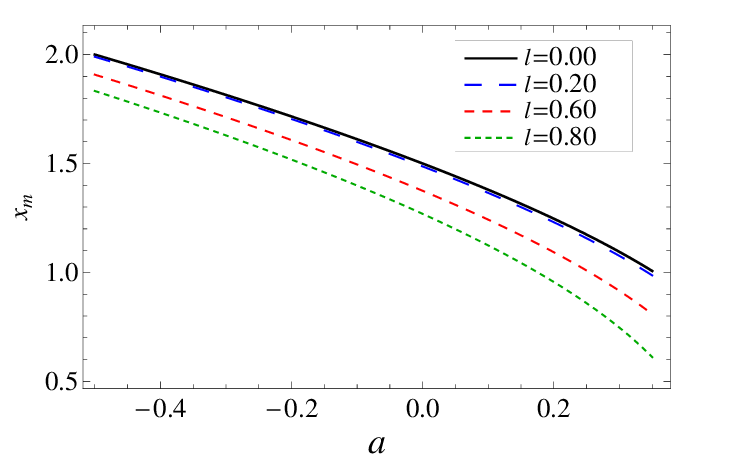}&
		     \includegraphics[scale=0.77]{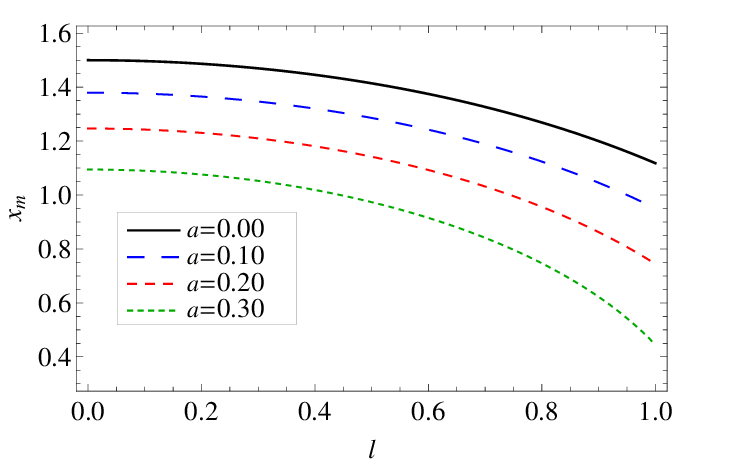}
		 \end{tabular}
	\end{centering}
	\caption{Plot showing radii of unstable circular photon orbit  for rotating Simpson-Visser black hole (dashed curve) in comparison to Kerr black hole (solid curve in the left plot) and spherical Simpson-Visser black hole (solid curve in the right plot). We  measure  the quantities $a$ and $l$ in units of Schwarzschild radius $2M$. }\label{plot3}	
\end{figure}

The Eq.~(\ref{ps1}) is solved numerically and the the radii of photon sphere for the black hole (\ref{NSR}) is depicted in Fig.~\ref{plot3} and is  smaller when compared with the Kerr black hole (cf. Fig.~\ref{plot3}). By assuming the closest distance $x_0$ very near to $x_m$, the deflection angle when expanded in the SDL takes the form  
\begin{eqnarray}\label{def}
\alpha_{D}(\theta)=-\bar{a} \log\Big(\frac{\theta D_{OL}}{u_m}-1\Big)+ \bar{b} + \mathcal{O}\left(u-u_m\right),
\end{eqnarray}
where $u$ is the impact parameter given by 
\begin{eqnarray}
u &=& L = \frac{\sqrt{(l^2+x_0^2)[a^2+l^2+x_0^2-\sqrt{l^2+x_0^2}}]-a}{\sqrt{l^2+x_0^2}-1},
\end{eqnarray}
and $u_m$ is the value of impact parameter at $x_0=x_m$ and is independent of the parameter $l$ (cf. Table~\ref{table1}).
Introducing  the  new variable $z=1-x_0/x$ to obtain the lensing coefficients as \cite{Bozza:2002af}
\begin{eqnarray}\label{bar}
\bar{a} &=& \frac{R(0,x_m)}{2\sqrt{{c_2}_m}}~~~\text{and} ~~~
\bar{b} = -\pi + b_R + \bar{a} \log\frac{c x_m^2 }{u_m^2},\\
b_R &=& \int_{0}^{1} [R(z,x_m)f(z,x_m)-R(0,x_m)f_0(z,x_m)]dz,\\
R(z,x_0) &=& \frac{2x^2}{x_0} \frac{\sqrt{B}\left(2A_0AL+A_0D\right)}{\sqrt{CA_0}\sqrt{4AC+D^2}},\\
f(z,x_0) &=& \frac{1}{\sqrt{A_0-A\frac{C_0}{C}+\frac{L}{C}\left(AD_0-A_0D\right)}}, ~~~~~~f_0(z,x_0) = \frac{1}{\sqrt{c_1 z + c_2 z^2}}.\\
u-u_m &=& c(x_0-x_m)^2
\end{eqnarray}

\begin{figure*}[htb!]
	\begin{centering}
		\begin{tabular}{c c}
		    \includegraphics[scale=0.75]{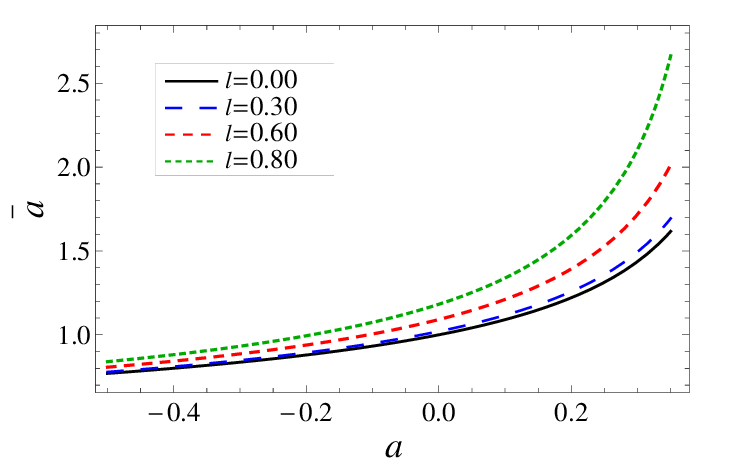}&
		    \includegraphics[scale=0.75]{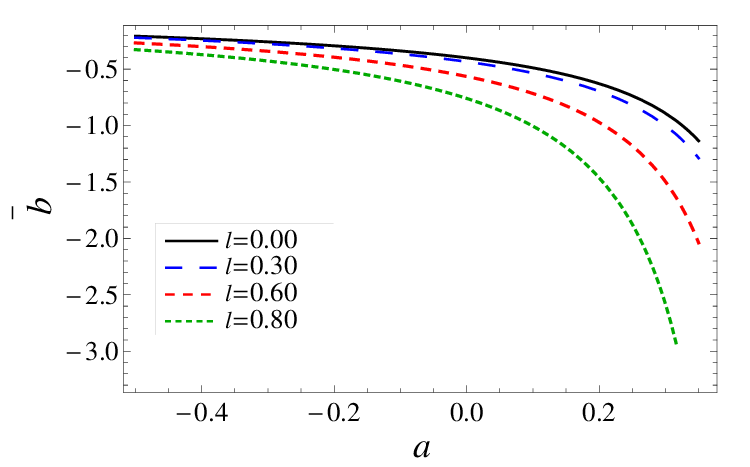}\\
		    \includegraphics[scale=0.75]{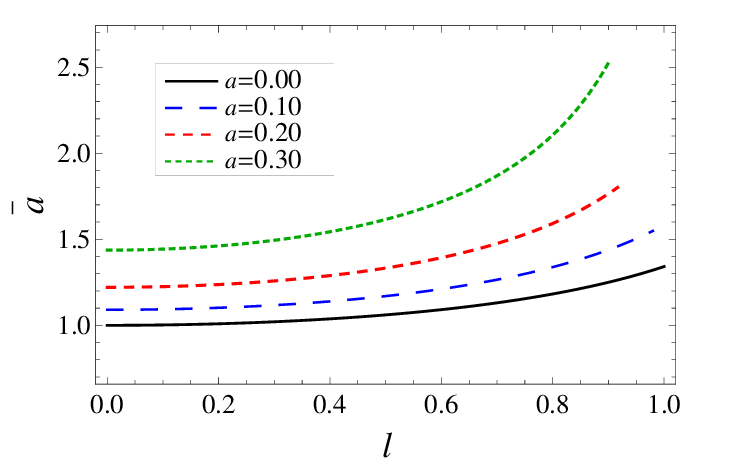}&
		    \includegraphics[scale=0.75]{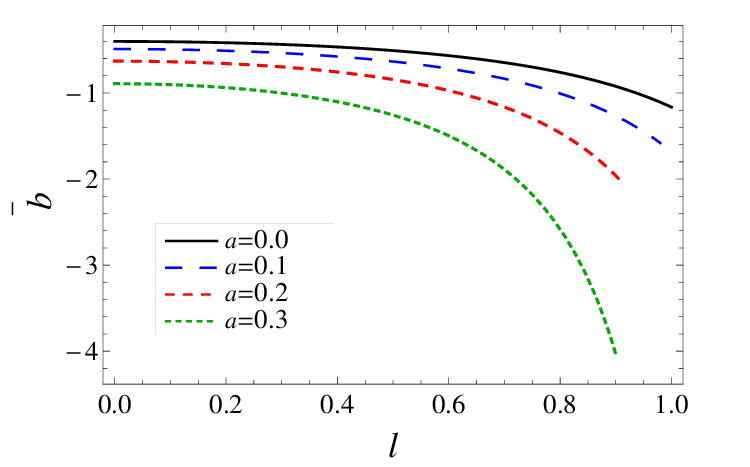}
			\end{tabular}
	\end{centering}
	\caption{Plot showing the behaviour of strong lensing coefficients $\bar{a}$ and $\bar{b}$. The solid black curve in the upper panel corresponds to Kerr black hole  and in the lower panel it corresponds to spherical Simpson-Visser black hole. We  measure  the quantities $a$ and $l$ in units of Schwarzschild radius $2M$.}\label{plot4}	
\end{figure*}

\begin{figure*}[htb!]
	\begin{centering}
		\begin{tabular}{cc}
		    \includegraphics[scale=0.75]{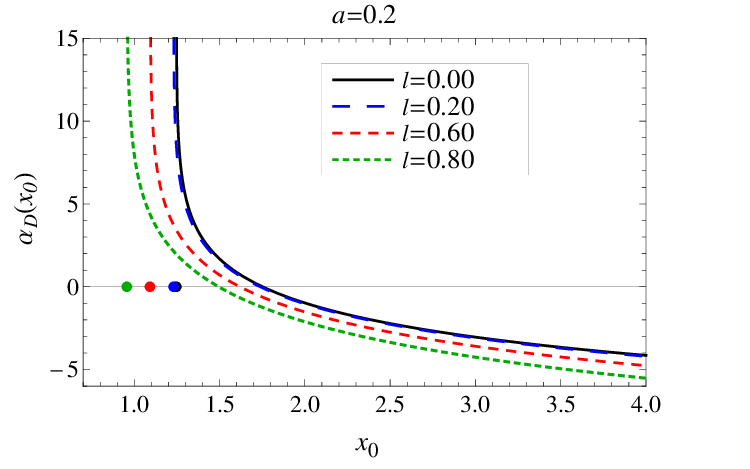}
		    \includegraphics[scale=0.75]{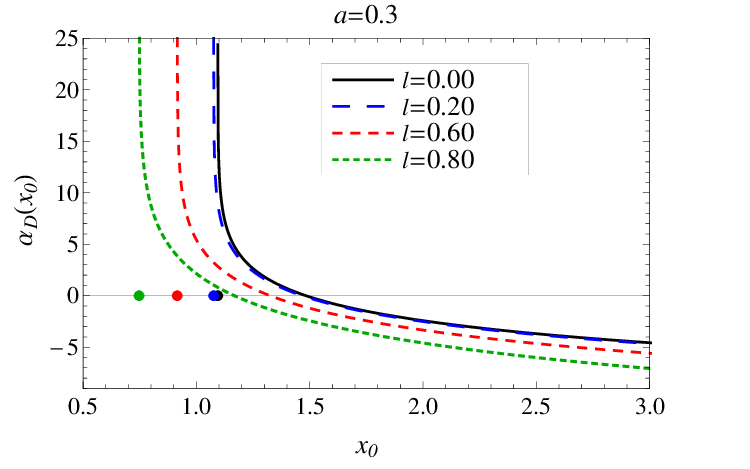}
			\end{tabular}
	\end{centering}
	\caption{Plot showing the variation of light deflection angle as a function of minimum distance $x_{0}$ for different values of $a$ and $l$. We  measure  the quantities $a$ and $l$ in units of Schwarzschild radius $2M$.} \label{plot5}		
\end{figure*}

\begin{figure*}[htb!]
	\begin{centering}
		\begin{tabular}{cc}
		    \includegraphics[scale=0.75]{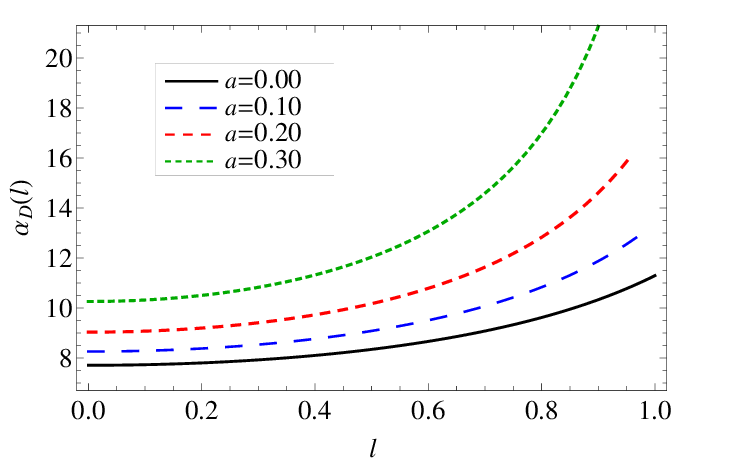}
			\end{tabular}
	\end{centering}
	\caption{Plot showing the variation of light deflection angle at $x_{0}=1.01 x_{m}$ as a function of parameter $l$.\label{plot6}}		
\end{figure*}

\begin{table*}[htb!]
\centering
\begin{tabular}{p{1cm}  p{1cm} p{1.6cm} p{1.6cm} p{1cm}}
\hline\hline
\multicolumn{2}{c}{}&
\multicolumn{2}{c}{Lensing Coefficients}&
\multicolumn{1}{c}{}\\
{$a$ } &  {$l$} & {$\bar{a}$}&{$\bar{b}$} & {$u_m/R_s$}\\ \hline
\hline
\\
\multirow{7}{*}{0.00} & 0.00  & 1.00000 & -0.400230 & 2.59808 \\          
                   & 0.25  & 1.01419 & -0.423912 & 2.59808\\      
                   & 0.50 & 1.06066 & -0.506783 & 2.59808\\      
                   & 0.75 & 1.15470 & -0.698106 & 2.59808\\
                   & 0.95  & 1.29219 & -1.03047 & 2.59808\\              
\hline 
\\
\multirow{7}{*}{0.15} & 0.00 & 1.14883 & -0.549965 & 2.28260\\      
                   & 0.25  & 1.17017 & -0.587359 & 2.28260\\      
                     & 0.50 & 1.24212 & -0.723681 & 2.28260\\           
                     & 0.75 & 1.39860 & -1.071530 & 2.28260\\
                     & 0.95 & 1.66148 & -1.79782 & 2.28260\\ 
                     
\hline
\\
\multirow{7}{*}{0.30}  & 0.00 & 1.43692& -0.891314& 1.91924\\
                   & 0.25  & 1.47594 & -0.965136 &1.91924\\      
                     & 0.50 & 1.61535 & -1.256170 & 1.91924\\           
                     & 0.75 &1.97301 & -2.180740 & 1.91924\\
                     & 0.90  &2.52527 &-4.037170 & 1.91924\\      
\hline
\\
\multirow{7}{*}{0.45}  & 0.00 & 2.58352 & -2.76735 & 1.42221\\
                   & 0.30  & 2.79948 & -3.26402 & 1.42221\\      
                     & 0.40 & 3.01084 & -3.79273 & 1.42221\\           
                     & 0.50 & 3.3693 & -4.7799 & 1.42221\\
                     & 0.60  &4.05135 & -6.94521 & 1.42221\\                            
\hline\hline
	\end{tabular}
	
\caption{Estimates for the strong lensing coefficients $\bar{a}$, $\bar{b}$ and the impact parameter $u_{m}/R_{s}$. We  measure  the quantities $a$ and $l$ in units of Schwarzschild radius $2M$.
\label{table1}  
}
\end{table*}   
Here the coefficients $c_1$, $c_2$ are obtained by the Taylor expansion of the term inside the square root of the $f(z, x_0)$. The behaviour of coefficients $\bar{a}$ and $\bar{b}$ is shown in Fig.~\ref{plot4}, and their behaviour is similar to the Kerr black hole ($l=0$) and spherical Simpson-Visser ($a=0$). The deflection angle for rotating Simpson-Visser black hole in SDL (cf. Fig~\ref{plot5} and Fig.~\ref{plot6}), like Kerr black hole, is more than $2\pi$, and multiple images of the source are formed. The photons that leave the source have to travel through different dispersive medium especially the environment around the black hole e.g the accretion disk, before it reaches the observer. The correction in  deflection angle (\ref{def}) can also be seen as coming from various environmental effects \cite{Aldi:2016ntn,Bozza:2010xqn}. It is just that we have not considered all possible sources of corrections such as black hole environment, e.g, accretion or plasma.

\subsection{Observables}
Next we shall  estimate the observables for the strong gravitational lensing by rotating Simpson-Visser black hole as in \cite{Islam:2021dyk,Ghosh:2020spb,Bozza:2002af,Bozza:2002zj}. Using the lens Eq.~(\ref{lenseq}), and following the condition where the source, lens and the observer are aligned, the angular seperation between the lens and the $n-$th image is given by~\cite{Bozza:2002af}  
\begin{eqnarray}\label{angpos}
\theta_n &=& \theta_n{^0} + \Delta\theta_n,
\end{eqnarray}
where 
\begin{eqnarray}
\theta_n{^0} &=& \frac{u_m}{D_{OL}}(1+e_n),\\
\Delta\theta_n &=& \frac{D_{OL}+D_{LS}}{D_{LS}}\frac{u_me_n}{\bar{a} D_{OL}}(\beta-\theta_n{^0}),\\
e_n &=& \text{exp}\left({\frac{\bar{b}}{\bar{a}}-\frac{2n\pi}{\bar{a}}}\right).
\end{eqnarray}

\begin{figure*}[htb!]
	\begin{centering}
		\begin{tabular}{cc}
		    \includegraphics[scale=0.7]{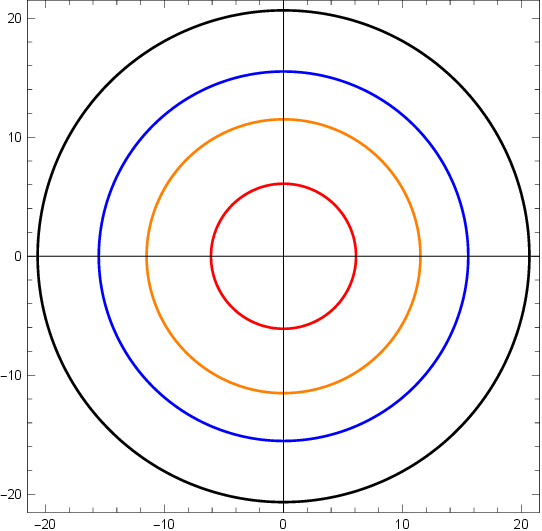}\hspace{1cm}
		    \includegraphics[scale=0.7]{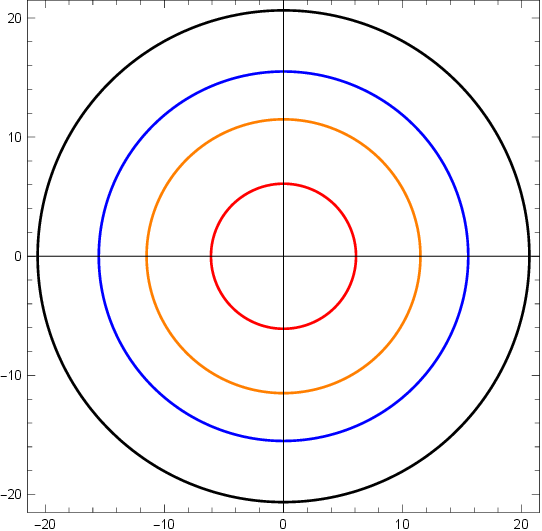}
			\end{tabular}
	\end{centering}
	\caption{The polar plot of $\theta_n^{E}$ (for $n=1$): The angular radius of the outermost Einstein ring. Here, we consider Kerr black hole ($a=0.25$ and $l=0$) (left) and rotating Simpson-Visser balck hole ($a=0.25$ and $l=0.7$) (right) as supermassive black holes  Sgr A* (black), M87* (blue), NGC 4649 (orange) and  NGC 1332 (red).}\label{plot8}		
\end{figure*}

Here $\theta_n{^0}$ is the angular position of the image when a photon
encircles complete $2n\pi$ and the second term in Eq.~(\ref{angpos}) is the extra term exceeding $2n\pi$ such that $\theta_n{^0} \gg \Delta\theta_n $ \cite{Bozza:2002zj}. For perfect alignment, i.e., when $\beta=0$, the lens deflects the light equally in all directions such that a ring-shaped image is produced, which are known as Einstein rings~\cite{Einstein:1936yew}. When the lens is exactly halfway between observer and source,
the angular radius of the Einstein rings can be obtained by solving Eq.~(\ref{angpos}) for  source, lens, and the observer being perfectly aligned  \cite{Muller:2008,Bozza:2002zj} as
\begin{eqnarray}\label{Ering3}
\theta_n^{E} &=& \frac{u_m}{D_{OL}}\left(1+e_n \right),
\end{eqnarray} 
where $n=1$ is the angular radius of the outermost Einstein ring (cf. Fig~\ref{plot8}) and the Einstein rings produced by the  Sgr A* are bigger when compared to other black holes \cite{Islam:2021dyk,Ghosh:2020spb}.   

The surface brightness is preserved in the deflection of the light but the appearance of the solid angle changes thus magnifying  the brightness of the images. For the $n-$loop images, the magnification is given by \cite{Bozza:2002af,Bozza:2002zj}
\begin{eqnarray}\label{mag}
\mu_n &=& \left|\frac{\beta}{\theta} \frac{\partial\beta}{\partial\theta}\right|^{-1}_{\theta_n^{0}} = \frac{1}{\beta} \Bigg[\frac{u_m}{D_{OL}}(1+e_n) \Bigg(\frac{D_{OL}+D_{LS}}{D_{LS}}\frac{u_me_n}{D_{OL} \bar{a}}  \Bigg)\Bigg].
\end{eqnarray}

\begin{table*}[htb!]
\resizebox{\textwidth}{!}{
 \begin{centering}	
	\begin{tabular}{p{0.8cm} p{1.5cm} p{2cm} p{1.5cm} p{2cm} p{1.5cm} p{2cm} p{1.5cm} p{1.2cm}}
\hline\hline
\multicolumn{1}{c}{}&
\multicolumn{2}{c}{Sgr A*}&
\multicolumn{2}{c}{M87*}& 
\multicolumn{2}{c}{NGC 4649}&
\multicolumn{2}{c}{NGC 1332}\\
{$a$ } & {$\theta_\infty$($\mu$as)}  & $\Delta T_{2,1}$(min) & {$\theta_\infty $($\mu$as)}  & $\Delta T_{2,1}$(hrs) & {$\theta_\infty $($\mu$as)}  &$\Delta T_{2,1}$(hrs) & {$\theta_\infty $($\mu$as)}  &$\Delta T_{2,1}$(hrs)\\ \hline
\hline
0.00    & 26.3299  & 11.4968  & 19.782  & 289.647 & 14.6615 & 210.328& 7.76719 & 113.185 \\
\hline 
0.15      & 23.1327  & 10.1008 &   17.38  & 254.477 & 12.8812 & 184.789 & 6.82406 & 99.441  \\
\hline
0.30 &  19.4504  & 8.4928 & 14.6133  & 213.968 & 10.8307 & 155.374 & 5.73777 & 83.612\\
\hline
0.45 &  14.4132  & 6.29295 & 10.8289  & 158.543 & 8.02585 & 115.127 & 4.2518 & 61.9538\\
		\hline\hline
	\end{tabular}
\end{centering}
}	
	\caption{Estimates for the observable $\theta_\infty$ and time delay $\Delta T_{2,1}$  for supermassive black holes at the center of nearby galaxies. These observables are independent of $l$.  
\label{table2}  
	}
\end{table*} 
In the case, $\beta \to 0$, the Eq.~(\ref{mag}) diverges, suggesting that the perfect alignment maximises the possibility of the detection of the images. The brightness of the first image is dominant over the other images, and in practice, if the 1-loop image can be distinguished from the rest packed images, we can have three characteristic observables~ \cite{Bozza:2002zj} as 
\begin{eqnarray}
\theta_\infty &=& \frac{u_m}{D_{OL}},\label{theta}\\
s &=& \theta_1-\theta_\infty \approx \theta_\infty ~\text{exp}\left({\frac{\bar{b}}{\bar{a}}-\frac{2\pi}{\bar{a}}}\right),\label{sep}\\
r_{\text{mag}} &=& \frac{\mu_1}{\sum{_{n=2}^\infty}\mu_n } \approx 
\frac{5 \pi}{\bar{a}~\text{log}(10)}\label{mag1}.
\end{eqnarray} 
In the above expression, $\theta_{\infty}$ is the angular position of the innermost image, $\theta_{1}$ is the angular position of the outermost image, $s$ is the angular separation between the first image and the inner packed images, $r_{\text{mag}}$ is the ratio of the flux of the first image and the all  other images. By observationally measuring the $s$,  $\theta_{\infty}$ and $r_{\text{mag}}$ and inverting the Eqs.~(\ref{theta}-\ref{mag1}), we can obtain the lensing the coefficients $\bar{a}$, $\bar{b}$, the minimum impact parameter as in $u_m$~\cite{Islam:2020xmy}. 

The time travelled by the light paths corresponding to the different images is not the same, and hence there is a time difference between the two images. The time delay, which is another important observable,   is defined as the time difference between the formation  of relativistic images considering the source with luminosity variations. The time for the photon to travel from the source and to the observer can be obtained using the null geodesic equation  as \cite{Bozza:2003cp}
\begin{eqnarray}
T &=& \int_{t_0}^{t_f} dt =  2\int_{x_{0}}^{\infty}\frac{dt}{dx}dx - \int_{D_{LS}}^{\infty}\frac{dt}{dx}dx -  \int_{D_{OL}}^{\infty}\frac{dt}{dx}dx,
\end{eqnarray}

where we have changed the integration variable from $t$ to $x$ and split the integral into approach and leaving phase and  by exploiting the symmetry between the two phases, we extend the integration to infinity.
Finally, if the time signal between the first and second image can be distinguished, the time delay $\Delta T_{2,1}$, when they are on the same side of the lens, can be approximated as  \cite{Bozza:2003cp}
\begin{eqnarray}
\Delta T_{2,1} \approx 2\pi\frac{\widetilde{R}(0,x_m)}{\bar{a}\sqrt{{c_2}_m}}=2\pi u_m,
\end{eqnarray}
where 
\begin{eqnarray}
\widetilde{R}(z,x_m) &=& \frac{2 x^2 \sqrt{B(x)A(x_0)}[C(x)-L D(x)]}{x_0 \sqrt{C(x)(D(x)^2+ 4 A(x)C(x))}}\left(1- \frac{1}{\sqrt{A(x_0)}f(z,x_0)}\right).
\end{eqnarray}
The essential requirement for measuring the time delay is that the source must be variable, which is not a restrictive requirement as variable stars are abundant in all galaxies. By implicitly assuming these variations, which will show up with a temporal phase in all images, it would be possible to measure the time difference between the relativistic images. From an observational perspective, the precise measurement of the time delay has an important advantage of dimensional measurement of the scale of the system, which can be used to get an accurate estimation of the distance of black hole and possibly to determine the Hubble parameter in cosmological context \cite{Refsdal:1964}.

\begin{table*}[htb!]
\resizebox{\textwidth}{!}{
 \begin{centering}	
	\begin{tabular}{p{0.8cm} p{0.8cm} p{1.8cm} p{1.8cm} p{1.8cm} p{1.8cm} p{1.8cm} p{1.8cm} p{1.8cm} p{1.8cm} p{1.8cm}}
\hline\hline
\multicolumn{2}{c}{}&
\multicolumn{2}{c}{Sgr A*}&
\multicolumn{2}{c}{M87*}& 
\multicolumn{2}{c}{NGC 4649}&
\multicolumn{2}{c}{NGC 1332}\\
{$a$ } & {$l$}& {$\theta_1$($\mu$as)} & {$s$ ($\mu$as)} & {$\theta_1$($\mu$as)}  & {$s$ ($\mu$as) } &  {$\theta_1$($\mu$as)}  & {$s$ ($\mu$as) } &  {$\theta_1$($\mu$as)}  & {$s$ ($\mu$as) } & {$r_{mag}$} \\ \hline
\hline

\multirow{7}{*}{0.00}  & 0.00 & 26.3628 & 0.0329517 & 19.8068  & 0.0247571 & 14.6799  & 0.0183488 & 7.77691  & 0.00972061 & 6.82188 \\

                         & 0.25 & 26.3652 &   0.0353455 &  19.8086  & 0.0265556 &  14.6812  & 0.0196818 &  7.77762 & 0.0104268 & 6.72647 \\      
                      
                        & 0.50  &  26.3735 &  0.0436767 &  19.8148  & 0.032815 & 14.6858 & 0.0243209 & 7.78008 & 0.0128844 & 6.43173\\                       
                     
                        & 0.75  & 26.3922  & 0.0623325 &  19.8289  & 0.0468314 &  14.6962 & 0.0347092 & 7.78558 & 0.0183878 & 5.90792\\
            
                       & 0.95  & 26.4216  & 0.091705 & 19.8509  & 0.0688993 & 14.7126 &  0.051065 & 7.79425 & 0.0270526 & 5.27931\\             
\hline 

\multirow{7}{*}{0.15}    & 0.00 & 23.1931  & 0.0604049 & 17.4254  & 0.0453831 & 12.9149 & 0.0336359 & 6.84188 & 0.0178192 & 5.93812\\      
                         & 0.25  & 23.198 &  0.0652086 &  17.429 & 0.0489922 & 12.9175 & 0.0363107 & 6.84329 & 0.0192362 & 5.82983 \\      
                      
                        & 0.50  &  23.2148 & 0.0821002 &  17.4417  & 0.0616831 &  12.9269  & 0.0457166 & 6.84828 & 0.0242192 & 5.49214\\                       
                     
                        & 0.75  & 23.2531 &  0.120348 &  17.4704 & 0.0904191 & 12.9482 & 0.0670144 & 6.85956 & 0.0355021 & 4.87764\\
          
                       & 0.95 & 23.3114 & 0.178621 & 17.5142 &  0.134201 & 12.9807 &0.0994633 & 6.87675 & 0.0526924 & 4.1059\\     
\hline

\multirow{7}{*}{0.30}    & 0.0  & 19.5823  & 0.131981 & 14.7125  & 0.0991593 & 10.9042 & 0.0734923 &  5.7767 & 0.0389338 & 4.74756\\
 
                         & 0.25  & 19.5936 &  0.143258  & 14.721 &  0.107632 & 10.9105 & 0.0797717 & 5.78003 & 0.0422604 & 4.62205 \\      
                      
                        & 0.50  & 19.6331  & 0.18278 & 14.7507  & 0.137326 & 10.9325  & 0.101779 & 5.79169 & 0.0539194 & 4.22317\\                       
                     
                        & 0.75  & 19.717 &  0.266607 & 14.8137  &  0.200306 & 10.9792  & 0.148458 & 5.81642  & 0.078648 & 3.45759\\
                      
                       & 0.90  & 19.777  & 0.326616 & 14.8587 & 0.245392 & 11.0126 & 0.181873 & 5.83412 & 0.0963504 & 2.70145 \\  
\hline

\multirow{7}{*}{0.45}    & 0.0  & 14.8471  & 0.433859 & 11.1548 & 0.325965 & 8.26744 & 0.24159 &  4.37982 & 0.127986 & 2.64054\\
 
                         & 0.30  & 14.8893 &  0.476064  & 11.1865 &  0.357674 & 8.29094 & 0.265092 & 4.39227 & 0.140437 & 2.43684 \\      
                      
                        & 0.40  & 14.9206  & 0.50742 & 11.21009 & 0.38123 & 8.3084  & 0.28255 & 4.40152 & 0.149689 & 2.26578\\                       
                     
                        & 0.50  & 14.9537 &  0.540453 & 11.2349  &  0.40605 & 8.32679  & 0.30094 & 4.41126  & 0.15943 & 2.02472\\
                      
                       & 0.60  & 14.9636  & 0.550440 & 11.2424 & 0.41355 & 8.33235 & 0.306507 & 4.41421 & 0.162377 & 1.15820 \\                        
		\hline\hline
	\end{tabular}
\end{centering}
}	
	\caption{Estimates for the lensing observables for supermassive black holes at the center of nearby galaxies for different values of $a$ and $l$. 
 We  measure  the quantities $a$ and $l$ in units of Schwarzschild radius $2M$.  }\label{table3}
\end{table*} 

\begin{figure*}[htb!]
	\begin{centering}
		\begin{tabular}{c c c c}
		    \includegraphics[scale=0.77]{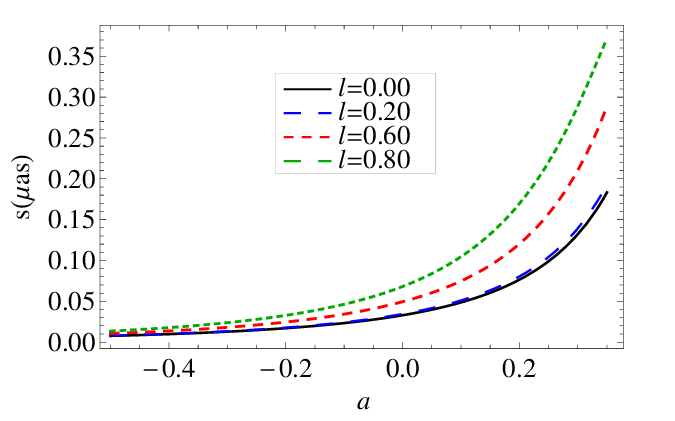}&
		    \includegraphics[scale=0.77]{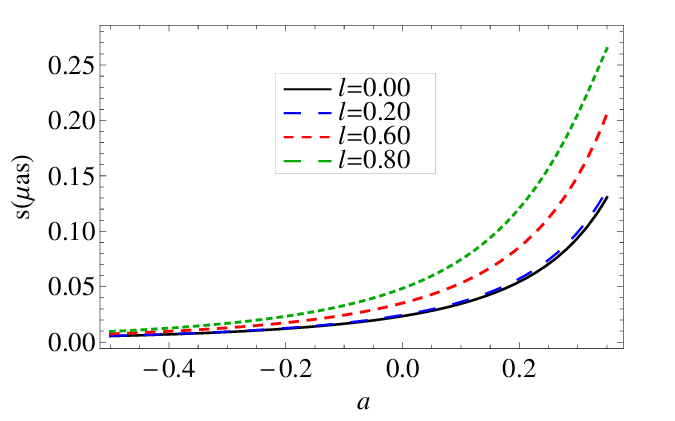}\\
		    \includegraphics[scale=0.77]{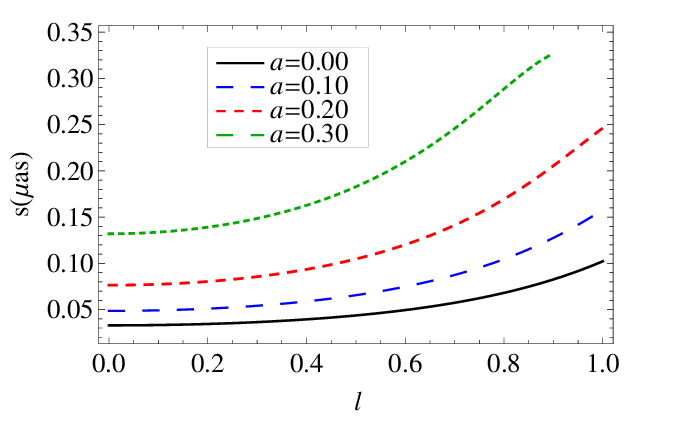}&
		    \includegraphics[scale=0.77]{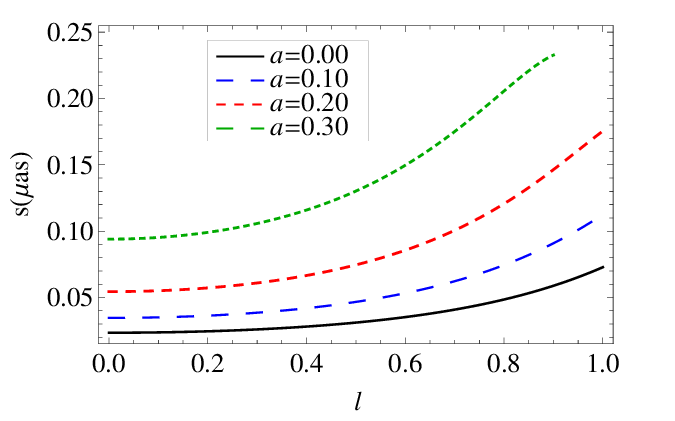}\\
			\includegraphics[scale=0.77]{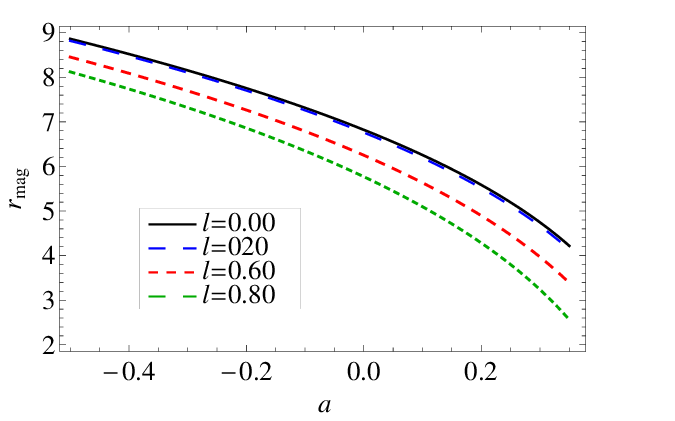}&
			\includegraphics[scale=0.77]{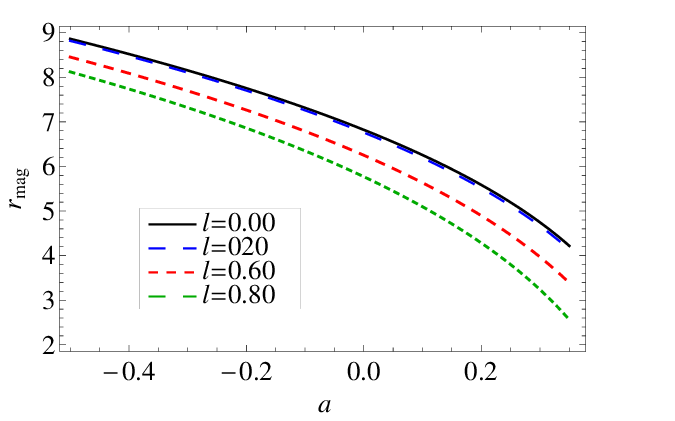}\\
			\includegraphics[scale=0.77]{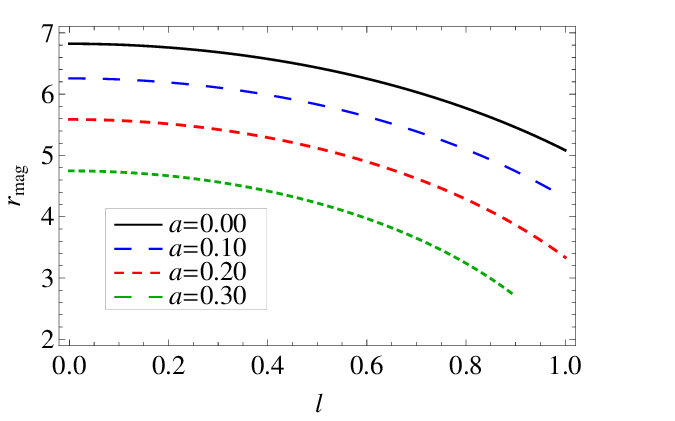}&
			\includegraphics[scale=0.77]{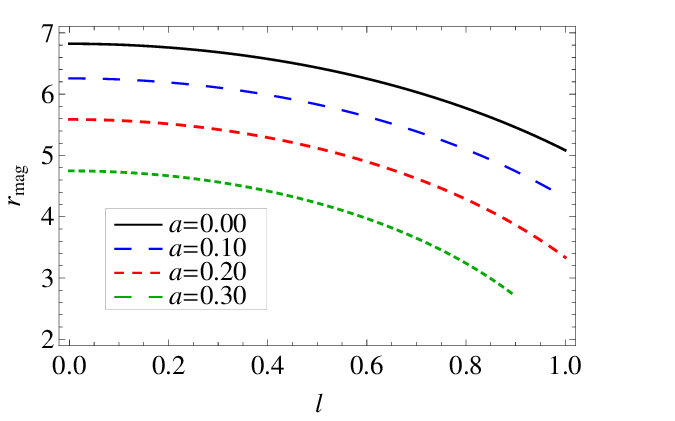}
			\end{tabular}
	\end{centering}
	\caption{Plot showing strong lensing observables  $s$, and $r_{mag}$ for the Sgr A*(left panel) and M87*(right panel) black holes. We  measure  the quantities $a$ and $l$ in units of Schwarzschild radius $2M$. \label{plot7}}
\end{figure*}

\section{Applications to supermassive black holes}\label{Sec5}
We apply the formalism in the previous section to estimate the observables in SDL by taking the  Sgr A*, M87*, NGC4649 and NGC1332  as the rotating Simpson-Visser black holes, and compare them with the corresponding quantities associated with Kerr black hole. The mass and distance from earth for Sgr A* \cite{Do:2019vob}  are $M \approx 4.3 \times 10^6 M_{\odot}$, $D_{OL}  \approx  8.35\text{kpc}$, for  M87* \cite{Akiyama:2019cqa} are  $M  \approx  6.5 \times 10^9 M_{\odot}$, $D_{OL}  \approx  16.8 \text{Mpc}$, for  NGC4649 \cite{Kormendy:2013} are  $M  \approx  4.72 \times 10^9 M_{\odot}$, $D_{OL}  \approx  16.46 \text{Mpc}$ and for  NGC1332 \cite{Kormendy:2013} are  $M  \approx  2.54 \times 10^9 M_{\odot}$, $D_{OL}  \approx  16.72 \text{Mpc}$, respectively. The  observables $s$ and $r_{mag}$ for rotating Simpson-Visser black holes for different values of $a$ and $l$ are depicted in Fig.~\ref{plot7}, whereas their values, along with the position of the first image $\theta_{1}$  are tabulated in Table~\ref{table3} and are compared with Schwarzschild black hole~($a=l=0$)  and Kerr black hole~($l=0$). The  observables $\theta_{\infty}$ and time delay $\Delta T_{2,1}$ do not depend on the parameter $l$ and are shown in Table~\ref{table2}. Considering hairy black holes as the lens, we observe that the angular position of first image images $\theta_{1}$ for Sgr A* and M87*, is in the range of $14.84~\mu\text{as} < \theta_{1}<26.36~\mu\text{as}$ and  $11.15~\mu\text{as} < \theta_{1}< 19.80~\mu\text{as}$, respectively. However, their deviations from the Kerr black hole are not more than  $0.1165~\mu$as and $0.087~\mu$as, respectively,  for $a = 0.45$. Although they deviate from the Kerr black hole, it is impossible to distinguish the rotating Simpson-Visser black holes from the Kerr black hole using the currently available observation facility of the EHT. Further, the angular separation $s$ between the first and inner most image due to the  rotating Simpson-Visser black hole for Sgr A* and M87* are in the range of $32.95~n\text{as} <s <433.8~n$as and   $24.75~n\text{as} <s< 413.55~n$as, respectively (cf. Table~\ref{table3}). The angular separation increases with both $l$ and $a$ suggesting that the image separation is larger than the corresponding Kerr black hole reaching a  deviation of  $116.5~n$as and $ 87.5~n$as for $a=0.45$, in case of Sgr A* and M87* respectively, which are beyond the threshold of the current  EHT observation (cf. Table~\ref{table4}). We may have to wait for the next generation event horizon telescope (ngEHT) for this purpose. The NGC4649 and NGC1332 show similar behaviour as Sgr A* and M87*, and to conserve space, we have not shown them.  If the angular resolution is large enough such that the first image can be resolved from the inner packed ones and it is possible to measure the brightness difference, we find that  the $r_{mag}$ of the images decreases with $l$ and $a$ for rotating Simpson-Visser black hole and thereby its images are less intense than that of Kerr black holes. The angular position and angular separation of the relativistic images, for given values of parameters $a$ and $l$, are larger for the Sgr A* black hole than the  M87*, NGC4649 and NGC1332 black holes. The time delay of the first image from that of the second image $\Delta T_{2,1}$,  for Sgr A*, M87*,  NGC4649 and NGC1332 respectively, can reach  $\sim11.49$~min, $\sim289.6$~hrs, $\sim210.3$~hrs  and $\sim113.2$~hrs (cf. Table~\ref{table2}). Thus, the time delay in Sgr A* is much shorter for observation and difficult for measurement. In the case of other black holes, the time delay  $\Delta T_{2,1}$ can reach in the order of a few hundred hours. These are sufficient time for astronomical measurements, provided we have enough angular resolution separating two relativistic images.

\begin{table*}[htb!]
\resizebox{\textwidth}{!}{
 \begin{centering}	
	\begin{tabular}{p{0.8cm} p{0.8cm} p{1.8cm} p{1.8cm} p{1.8cm} p{1.8cm} p{1.8cm} p{1.8cm} p{1.8cm} p{1.8cm} p{1.8cm}}
\hline\hline
\multicolumn{2}{c}{}&
\multicolumn{2}{c}{Sgr A*}&
\multicolumn{2}{c}{M87*}& 
\multicolumn{2}{c}{NGC 4649}&
\multicolumn{2}{c}{NGC 1332}\\
{$a$ } & {$l$}& {$\Delta\theta_1$($\mu$as)} & {$\Delta s$ ($\mu$as)} & {$\Delta\theta_1$($\mu$as)}  & {$\Delta s$ ($\mu$as) } &  {$\Delta \theta_1$($\mu$as)}  & {$\Delta s$ ($\mu$as) } &  {$\Delta \theta_1$($\mu$as)}  & {$\Delta s$ ($\mu$as) } & {$\Delta r_{mag}$} \\ \hline
\hline

\multirow{7}{*}{0.30} & 0.25  &  0.0112769 & 0.0112769  &  0.00803363 & 0.00803363 &    0.00627941 & 0.00627941 & 0.0149856 & 0.0149856 &  -0.125517 \\      
                      
                        & 0.50  & 0.0507993 & 0.0507993  &0.0361893 & 0.0361893 &0.028287 & 0.028287 &  0.0149856 & 0.0149856 & -0.524392\\                       
                     
                        & 0.75  & 0.134626 & 0.134626 & 0.0959075 & 0.0959075 & 0.0749652 & 0.0397141 & 0.0397141 & 0.078648 & -1.28997\\
                      
                       & 0.90  &  0.194635 & 0.194635 &  0.138658 & 0.138658 &  0.108381 &  0.0574166 & 0.0574166 & 0.0963504 & -2.04612\\
  	\hline                     
\multirow{7}{*}{0.45} & 0.30  &  0.0422 & 0.042205  &  0.0317 & 0.0317098 &    0.0235 & 0.0235018 & 0.0124 & 0.0124505 &  -0.2037 \\      
                      
                        & 0.40  & 0.0735 & 0.073560  & 0.0552 & 0.0552727 & 0.0409 & 0.0409655 &  0.02170 & 0.0217022 & -0.3747\\                       
                     
                        & 0.50  & 0.1065 & 0.10659 &  0.08008 & 0.0800856 & 0.05935 & 0.0593557 & 0.03144 & 0.0314447 & -0.615823\\
                      
                       & 0.60  &  0.1165 & 0.116582 &  0.08758 & 0.0875896 &  0.06491 &  0.0649173 & 0.03439 & 0.0343911 & -0.956684 \\  
		\hline\hline
	\end{tabular}
\end{centering}
}	
	\caption{Deviation of the lensing observables of rotating Simpson Visser
	black holes from Kerr black hole for supermassive black holes at the center of nearby galaxies for  $a=0.3$ and $a=0.45$. For $a=0.3$  the rotating Simpson-Visser black hole admits  both Cauchy and  event horizon for $0.0<l_{cb}<0.1$ and only event horizon for $0.1<l_{c}<0.9$.  For $a=0.45$,  both Cauchy and  event horizon exist for $0.0<l_{cb}<0.282055$  and only event horizon is present for $0.282055<l_{c}<0.717945$. Here $\Delta(X)=X_{RSV}-X_{Kerr}$ 
\label{table4}  
	}
\end{table*} 

\section{Lensing in non-equatorial plane}\label{Sec7}
\begin{figure*}[htb!]
	\begin{centering}
		\begin{tabular}{cc}
		    \includegraphics[scale=0.25]{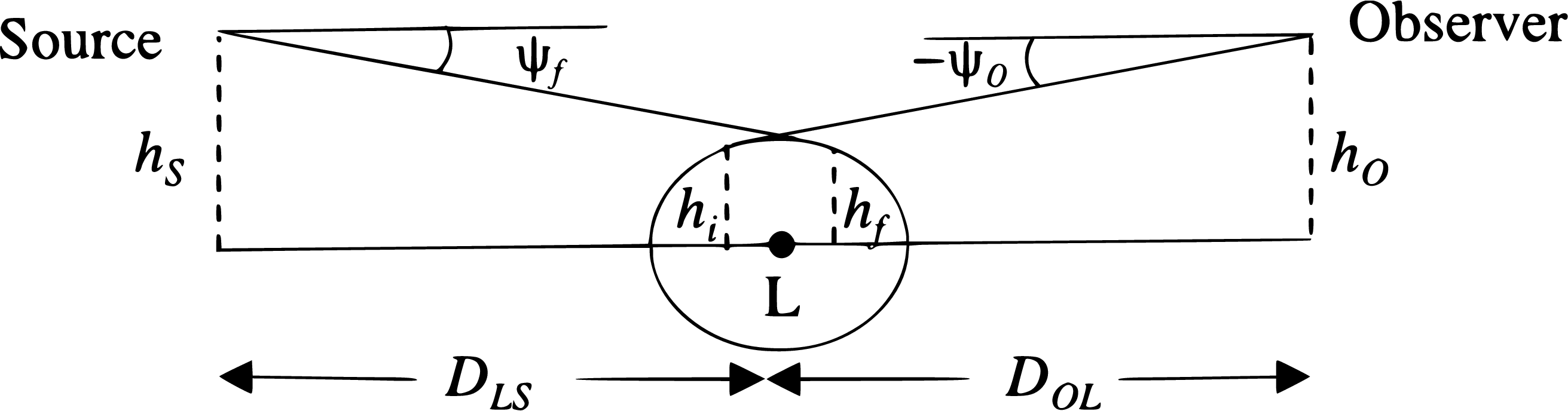}
			\end{tabular}
	\end{centering}
	\caption{The lensing configuration of a non equatorial observer projected on a vertical plane. }\label{fig9b}		
\end{figure*}

In the discussion in the previous section on the equatorial plane, we write just a one-dimensional lens equation. However, for practical purposes, e.g., to discuss the caustic structure and find the magnification of the image, it is necessary to consider the two-dimensional lens equation. Thus, it is pertinent to investigate the non-equatorial lensing by rotating Simpson-Visser black hole. We shall analyse the ray trajectories close to the equatorial plane \cite{Bozza:2002af} of the rotating Simpson-Visser black hole. An additional polar coordinate $\psi$ is required to derive the two-dimensional lens equation so that $\psi=\pi/2-\theta$. Further, the light ray trajectory is identified by three parameters, viz, $\psi_0$, $u$ and $h$ (cf. Fig~\ref{fig9b}). $\psi_0$ is the declination that the incoming trajectory makes with the Equatorial plane, $u$ is the impact parameter of the projection of the light ray on the Equatorial plane, and lastly, the distance between that point of the projection closer to the black hole and the trajectory is denoted by $h$.

For an observer located at ($\theta_0, x_0$), two celestial coordinates $\zeta_1$ and $\zeta_2$ can be defined for an image in the Boyer-Lindquist system. Precisely, $\zeta_1$ is the perpendicular distance of the image from the axis of symmetry and $\zeta_2$ is the perpendicular distance of the image from its projection on the equatorial plane. The coordinates $\zeta_1$ and $\zeta_2$  are  obtained using the geodesics equations as follows \cite{Chander:1992pc,Bozza:2006nm}
\begin{eqnarray}
\zeta_1 &=& x_0^{2} \sin \theta_0  \frac{d\phi}{dx} \bigg\vert_{x_0\to \infty} = \frac{L}{\sin\theta_0} \nonumber\\
\zeta_2 &=& x_0^2  \frac{d\phi}{dx} \bigg\vert_{x_0\to \infty} = h \sin\theta_0
\end{eqnarray}
Taking into consideration that $\theta_0=\pi/2 - \psi_0$ and $\zeta_1 = u$, we can relate the constants of motion $L$ and  $Q$ to the initial parameters characterising the trajectory, i.e., $u$ and $\psi_0$  as \cite{Chander:1992pc,Bozza:2002af}
\begin{eqnarray} \label{lq}
L &=& u \cos\psi_0 \nonumber\\
Q &=& h^2 \cos^2 \psi_0 + (u^2-a^2) \sin^2\psi_0
\end{eqnarray}
Here, $L$ and $Q$ are the impact parameters in the non-equatorial planes.  The unstable orbit with constant $x$ in case of non equatorial orbits is given when $L$ and $Q$ lie on the critical locus. The non equatorial orbits are characterised by $Q_m>0$, with the value of $x_m$ in between $x_m^+$ and $x_m^-$, which, for the equatorial observer  are the  radius of the circular orbit for  prograde and retrograde  respectively.

\begin{figure}[htb!]
	\begin{centering}
		\begin{tabular}{cc}
		    \includegraphics[scale=0.75]{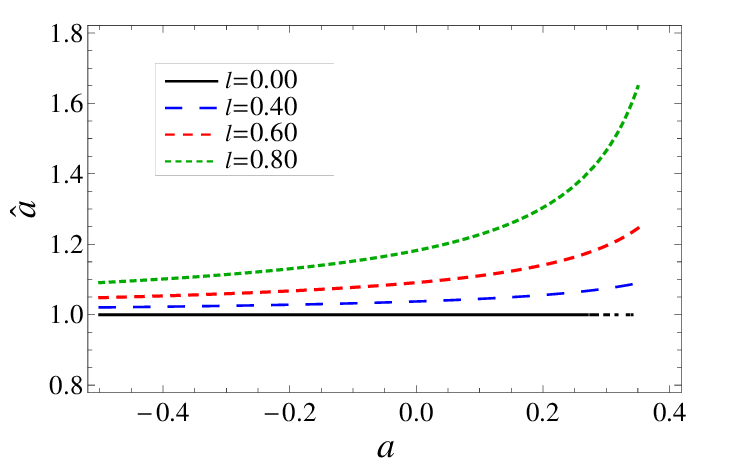}&
		    \includegraphics[scale=0.75]{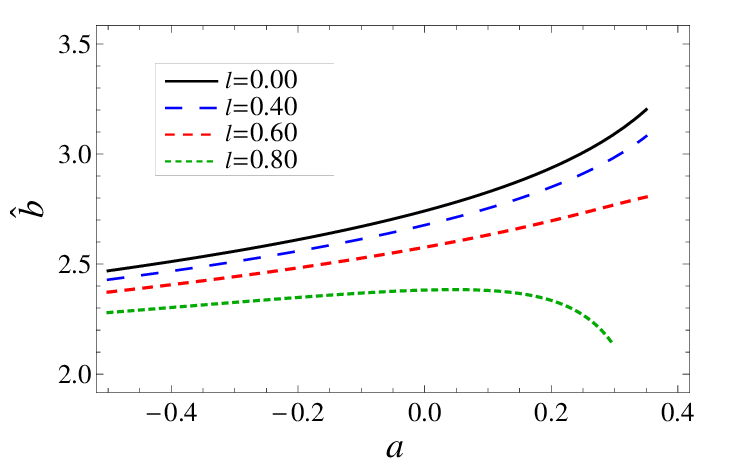}\\
		 \end{tabular}
\end{centering}
\caption{Plot showing the behaviour of strong lensing coefficients $\hat{a}$ and $\hat{b}$. The solid black curve  corresponds to Kerr black hole ($l=0$).  We  measure  the quantities $a$ and $l$ in units of Schwarzschild radius $2M$.}\label{plot9}		
\end{figure}

Following Bozza~\cite{Bozza:2002af}, we restrict our calculation to small inward declination $\psi_0$ and small height. Accordingly, we write the Eq.~(\ref{lq}) up to first order terms as 
\begin{eqnarray} \label{lq1}
L \approx u  ~~~~\text{and}~~~
Q \approx h^2 + (u^2-a^2)\psi^2{_0}
\end{eqnarray} 
 
The evolution of the $\psi$ as a function of azimuthal angle $\phi$ during the motion of the photons around the  rotating Simpson-Visser black hole is simply given by 
\begin{eqnarray} \label{psi}
\frac{d\psi}{d\phi} &=& \pm \omega(\phi) \sqrt{\bar{\psi}^2-\psi^2}
\end{eqnarray}
where
\begin{eqnarray}
\bar{\psi}^2 &=& \frac{h^2}{\bar{u}^2}+\psi_0^2\\
\omega(\phi) &=& \bar{u}\frac{x^2+l^2+a^2-\sqrt{x^2+l^2}}{u(x^2+l^2-\sqrt{x^2+l^2}) + a\sqrt{x^2+l^2}}\\
\bar{u} &=& \sqrt{u^2-a^2}
\end{eqnarray}
The parameter $\omega(\phi)$ depends on the deviation $l$ and spin $a$ and reduces to the Kerr black hole case for $l=0$. Since $x$ depends on $\psi$, the solution of the Eq.~(\ref{psi}) is of the form 
\begin{eqnarray}
\psi(\phi) &=& \bar{\psi}\cos(\bar{\phi} + \phi_0)
\end{eqnarray}
with
\begin{eqnarray}
\bar{\phi} = \int_0^{\phi_f}\omega(\phi')d\phi'.
\end{eqnarray}
where $\phi_f$ is the  total azimuthal change experienced by the photon and $\phi_0$ is a constant. The integral in this case can be simplied to  
\begin{eqnarray}\label{def2}
\bar{\phi}_f &=& 2\int_{x_0}^{\infty}\omega(x)\frac{d\phi}{dx}dx = \int_{0}^{1}\omega(x)R(z,x_0)f(z,x_0)dz.
\end{eqnarray}

\begin{figure}[htb!]
	\begin{centering}
		\begin{tabular}{c c}
		    \includegraphics[scale=0.85]{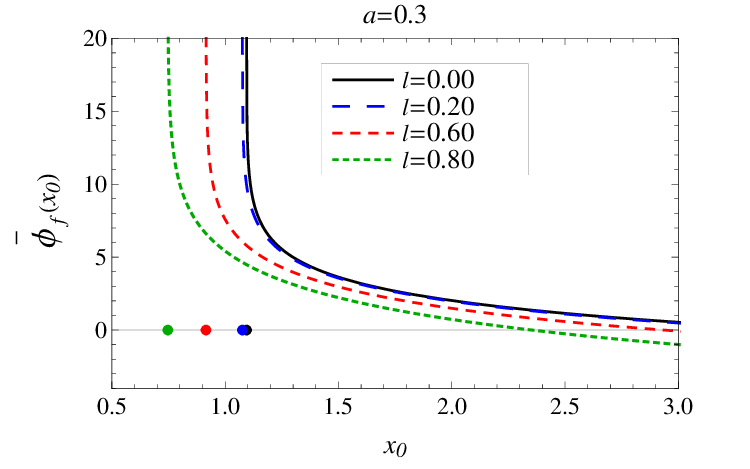}&
		    \end{tabular}
	\end{centering}
	\caption{Plot showing the variation of light  $\bar{\phi_f}$ as a function of minimum distance $x_{0}$.  We  measure  the quantities $a$ and $l$ in units of Schwarzschild radius $2M$.}\label{plot10}		
\end{figure} 

The term $\omega(x)$ is free of singularities and can be absorbed into the  function $R(z,x_0)$, such that, we can define the quantity $R_{\omega}(z,x_0) = \omega(x) R(z,x_0)$ which is regular for all values of $x_0$. Following the same technique we have used for equatorial lensing, Eq.~(\ref{def2}) can be expressed in terms of lensing coefficients as
\begin{eqnarray}
\bar{\phi}_f &=& -\hat{a} \log\Big(\frac{u}{u_m}-1\Big)+ \hat{b}\\\label{def3}
\hat{a} &=& \frac{R_{\omega}(0,x_m)}{2\sqrt{{c_2}_m}}~~~\text{and} ~~~
\hat{b} = \hat{b}_R + \hat{a} \log\frac{c x_m^2 }{u_m^2},\\
\hat{b}_R &=& \int_{0}^{1} [R_{\omega}(z,x_m)f(z,x_m)-R_{\omega}(0,x_m)f_0(z,x_m)]dz.
\end{eqnarray}

The lensing coefficients $\hat{a}$ and $\hat{b}$ are evaluated at $x_m$ and depicted in Fig.~\ref{plot9}.  Interestingly, for the case of Kerr black hole we have $\hat{a}=1$ for all values spin  \cite{Bozza:2002af}. However, for $l \ne 0$, $\hat{a}$ shows similar behaviour as $\bar{a}$. On the other  $\hat{b}$ tends to increase with $a$ for smaller values of $l$, which is contrary to the behaviour of $\bar{b}$. Variation of $\bar{\phi_f}$ is presented in Fig.~\ref{plot10} and as can be seen it tends to zero at slightly higher values of $x_0$ than $\alpha_{D}(\theta)$. The factor $\omega(x) <1$ for prograde photons and these photons lag the initial declination after each loop while as for retrograde photons $\omega(x)>1$ and  photons tend to  lead the initial declination after each loop. Further, we consider the lens geometry by setting the source at some height $h_s$  and the observer at height $h_o$ on the equatorial plane. By assuming that $u<<(h_o,h_s)<<D_{OL},D_{LS}$, we can write the polar lens equation as

\begin{eqnarray}\label{height}
h_s &=& h_o\left( \frac{D_{OL}}{\bar{u}}S-C\right)-\psi_0\left[(D_{OL}+D_{LS})C-\frac{D_{OL}D_{LS}}{\bar{u}}S\right],
\end{eqnarray}
where $S=\sin \bar{\phi}_f$ and $C=\cos \bar{\phi}_f$. Solving  Eq.~(\ref{height}) for $\psi_0$, the inclination and the height  for the $n-$th image will be 
\begin{eqnarray}\label{height1}
\psi_{0,n} &=& \frac{(h_s+h_o\cos\bar{\phi}_{f,n})\bar{u}-h_o D_{LS}~\sin \bar{\phi}_{f,n}}{D_{OL}D_{LS}~\sin \bar{\phi}_{f,n}-\bar{u}(D_{OL}+D_{LS})\cos \bar{\phi}_{f,n}} \nonumber\\
h_n &=& \frac{\bar{u}(h_s D_{OL}-h_o D_{LS}\cos\bar{\phi}_{f,n})}{D_{OL}D_{LS}~\sin \bar{\phi}_{f,n}-\bar{u}(D_{OL}+D_{LS})\cos \bar{\phi}_{f,n}}.
\end{eqnarray}
The quantity $\bar{\phi}_{f,n}$ is the phase of the $n$-th image, which is defined by the Eq.~(\ref{def3}) and can be deduced once we obtain the impact angle $\theta_n$ of $n$-th image using equatorial lens equation. The denominators of terms $\psi_{0,n}$ and $h_n$ can vanish in the neighbourhood of $\bar{\phi}_{n}=k\pi$, which defines the position of caustic points. 
\section{Conclusions}\label{Sec6}
The spacetime singularities in classical general relativity are inevitable, which are also predicted by the celebrated singularity theorems.  However, it is a general belief that singularities do not exist in nature, and they are the limitations of general relativity.  In the absence of well-defined quantum gravity, models of regular black holes have been studied.  Therefore, it is instructive to explore the gravitational lensing to assess the dependence of observables on the  parameter $l$  and compare the results with those for the Kerr black holes. 

With this motivation, we have analyzed the strong gravitational lensing of light due to rotating Simpson-Visser black holes, which besides the mass $M$ and angular momentum $a$ has an additional deviation parameter $l$. We have examined the effects of the black hole  parameter $l$ on the light deflection angle $\alpha_{D}(\theta)$ and lensing observables $ \theta_{\infty} $, $s$, $r_{\text{mag}}$,  $u_{m}$, in the strong field observation, due to rotating Simpson-Visser black holes and compare them to the Kerr black holes. We have analytically derived the first order geodesic equations in the background of Simpson-Visser black holes. The observer and source's geometrical relation has been expressed using an asymptotic lens equation that we combined with geodesic equations to obtain the strong lensing coefficients $\bar{a}$, $\bar{b}$ and deflection angle. We have numerically calculated the strong lensing coefficients and lensing observables as functions of $l$ and $a$ for the relativistic images. we find that $\bar{a}$ increases whereas $\bar{b}$ and deflection angle $\alpha_D$ decrease with increasing $l$. $u_m$ depends only on the spin of the black hole and is independent of the black hole parameter $l$. We have calculated the position, separation and  relative magnification  of these images for supermassive black holes, namely, Sgr A*, M87, NGC 4649 and NGC 1332. Time delay $\Delta T_{2,1}$ of the first and second image when they are on the same side of the source  has also been obtained.  The observables $\theta_{\infty}$ and $\Delta T_{2,1}$  do not depend  of $l$. Further, the relative magnification of the first and innermost images is independent of the black hole mass or its distance from the observer. Interestingly, it turns out that the separation between the two relativistic images for Simpson-Visser black holes is greater than that for the Kerr black holes.  

The shadow of M87* by EHT is consistent with the Kerr black hole's image predicted by the GR \cite{Akiyama:2019cqa,Akiyama:2019bqs}. Some rotating black holes, arising in modified theories of gravity, are similar to Kerr, with subtle differences on an astrophysical scale. The reported observations do not allow us to rule out these scenarios. Motivated by this also, we investigated the rotating Simpson-Visser black holes, in the strong deflection limit.  Although, our results constrain the rotating Simpson-Visser black holes parameter $l$ from the gravitational lensing analysis (cf. Figs.~\ref{plot5}, \ref{plot6} and \ref{plot7}) and horizon structure (cf. Figs.~\ref{plot1} and \ref{plota}).  However, the departure produced by the parameter $l$  for lensing observables by taking the rotating Simpson-Visser black holes as the supermassive black holes, namely, Sgr A*, M87, NGC 4649 and NGC 1332, from those for Kerr black hole for $0<l <0.6$ ($a=0.45$) are $\mathcal{O}(\mu$as). Hence,  it is difficult to distinguish the regular black holes from the Kerr black hole, at least from the present resolution of astronomical observations like the EHT.  Therefore our results, in principle, could provide a possibility to test how rotating Simpson-Visser black holes deviates from the Kerr black hole by future astronomical observations like ngEHT. Our results are the generalization of previous discussions, on the Kerr black holes, in a more general setting.   

We have also investigated the non-equatorial (quasi-equatorial) lensing in strong field limit by the rotating Simpson-Visser black hole at very small declination and calculated the deflection angle. An interesting behaviour of $\hat{a}$ is that it goes to unity for all values of spin $a$ for the Kerr black hole. $\hat{b}$, on the other hand, increases with $a$ for small values of $l$.  The caustic structure not limited to the equatorial plane of rotating Simpson-Visser black holes can help us better understand the phenomenology of the black holes at the centre of our and nearby galaxies, which will be an exciting project.    

Many interesting avenues are amenable for future work from the rotating Simpson-Visser solutions; it will be intriguing to analyze the relationship between the null geodesics and thermodynamic phase transition in AdS background in the context photon sphere. One should also see the possibility of generalization of these results to investigate the higher curvature effect in the gravitational lensing. Finally, the possibility of a different conception of these result to gravitational lensing for the solution (\ref{metric1}), that describes a wormhole with a throat located at $r = 0$, is an interesting problem for future research.  In particular, our results, in the limit $l \rightarrow 0$, reduced exactly  to  \emph{vis-$\grave{a}$-vis}  the Kerr black hole results.

It will be intriguing to investigate gravitational lensing in a plasma or accretion environment. Further,  the strong field gravitational lensing may open fascinating perspectives for testing modified theories of gravity and estimating the parameters associated with the supermassive black holes.

\section{Acknowledgments} 
S.G.G. and  S.U.I.  would like to thank SERB-DST for the ASEAN project IMRC/AISTDF/CRD/2018/000042. J.K. would like to thank CSIR for providing JRF. We would like to thank the anonymous reviewer for their valuable comments and suggestions.

 \end{document}